\documentclass[amssymb,twocolumn,
aps,floatfix,showpacs]{revtex4}
\usepackage{graphicx}

\begin{document}
\title{The influence of multiple ionization thresholds on harmonic generation: Ar$^+$ }
\author{A. C. Brown and H. W. van der Hart}
\affiliation{Centre for Theoretical Atomic, Molecular and Optical Physics,
Queen's University Belfast, Belfast, BT7 1NN, UK.}
\date{}

\begin{abstract} 
  We apply time-dependent $R$-matrix theory to investigate harmonic generation
  from ground state Ar$^+$ with $M=0$ at a wavelength of 390-nm. Contributions
  associated with the different $3s^23p^4$ ionization thresholds are assessed,
  including the interference between these. The dominant contribution originates
  from the second ionization threshold, $3s^23p^4$ $^1D$. Changes to the
  harmonic yields arising from the higher $3s3p^5$ thresholds are also assessed.
  We further confirm that Ar$^+$ has a higher harmonic yield than He for the
  same laser pulse, despite having a higher ionization threshold.  
\end{abstract}

\pacs{32.80.Rm, 31.15.A-, 42.65.Ky}
\maketitle

\section{Introduction} 

Harmonic generation (HG) has been at the centre of atomic and molecular
physics for over 20 years \cite{attosecond_review}. Not only is it the
means of producing the ultrashort laser pulses which drive attosecond physics
\cite{attosecond_pulse_train}, but it has been adopted as an important measurement tool for some
of the most fundamental physical atomic and molecular processes. HG has facilitated
imaging molecular dynamics \cite{proton_dynamics_molecules}, obtaining detailed information
about molecular orbitals \cite{corkum_review} and
has been employed to highlight the importance of electron correlation and multielectron
dynamics in a variety of systems on ultrashort timescales
\cite{multielectron_molecules,multielectron_atoms,brown_prl}.

Generally, HG is described in terms of a three-step model \cite{corkum1993}:
a laser field causes tunnel ionization of an electron, which is
subsequently accelerated in the field. As the electric field changes direction,
the electron is driven back towards its parent ion, and can be recaptured,
emitting a high-energy photon.  A semiclassical model of this
process has proven highly successful in explaining
experimental phenomena, particularly in noble-gas atoms \cite{lewenstein}.

The three-step model describes the main physics leading to HG, but the model
cannot be applied directly to all atomic or molecular systems. It assumes that
tunnel ionization leaves the residual ion in the ground state.  This is
appropriate for noble gas atoms, but may not be so for systems in which the
residual ionic state lies close to other ionic states. Examples of such systems
are primarily found in molecules such as N$_2$. It is less appreciated that they
can also be found in atomic systems, eg. Ar$^+$.  For these systems,
electron-emission channels associated with different ionic states contribute to
HG, and interference between these channels can be of significance.  Since HG
occurs within a laser field cycle, this interference provides information about
ultrafast dynamics. To develop understanding of how this information can be
extracted, accurate theoretical data is of great benefit. To obtain such data,
it is imperative to apply theoretical methods capable of including multiple
channels associated with different thresholds and the interactions between these
channels: recent studies on the Cooper minimum in Ar have shown that the
inclusion of multichannel effects can alter the harmonic yield by as much as two
orders of magnitude \cite{multichannel_hhg}. 

We have recently developed capability within time-dependent $R$-matrix theory
(TDRM) \cite{tdrm} to determine harmonic spectra for general multielectron atoms
\cite{brown_prl,brown_helium}.  TDRM is well-suited to study ultrafast
multielectron dynamics, as demonstrated in a study of C$^+$ \cite{collect_c+1},
and multichannel interferences in HG, as evidenced in a study of resonant
enhancement of the 5th harmonic in Ar \cite{brown_prl}.  In order to study the
interplay of multiple channels associated with low-lying thresholds, we apply
TDRM to investigate HG from Ar$^+$ in a $4\times 10^{14}$ Wcm$^{-2}$, 390-nm
laser field. This intensity/wavelength regime can lead to significant
non-perturbative changes in the atomic structure.  Therefore, perturbative
methods are not suitable for addressing an ion in these fields, even though the
interaction can be characterised as a multiphoton process. Given the small
energy gap between the relevant thresholds, we would expect that the general
multichannel effects seen for 390-nm would still be important at 800-nm, even
though the fine detail may differ.   

Ar$^+$ is an ion well-suited to investigations on multiple thresholds: the
lowest three ionization thresholds, $3s^23p^4$ $^3P$, $^1D$ and $^1S$, are
separated from each other by about 2 eV (Tab. \ref{tab:nist-data}). We can also
investigate the effects of higher-lying thresholds by including the $3s3p^5$
thresholds. Ar$^+$ thus allows the investigation of interference effects arising
from the interplay between channels associated with well separated, as well as
closely spaced, thresholds.

\begin{table}[t]
  \caption{ Energies of the five ionization thresholds of Ar$^+$ included in the present work 
with respect to the Ar$^{2+}$ ground state, and compared to
literature values \cite{argon+_nist}.
\label{tab:nist-data}}  
   \begin{tabular*}{\columnwidth}{@{\extracolsep{\fill}}cccc}
	  \hline
	  \hline
	  Configuration  & Term & Energy (Lit.) & Energy (TDRM) \\
	   	  &     &    {\footnotesize eV }        &      {\footnotesize eV}      \\
	  \hline 
	  $3s^23p^4$ & $^3P\phantom{^o}$    & \phantom{0}0.00  & \phantom{0}0.00\\
			   	 & $^1D\phantom{^o}$    & \phantom{0}1.67  & \phantom{0}2.04\\
					 & $^1S\phantom{^o}$    & \phantom{0}4.06  & \phantom{0}3.85\\
	  $3s3p^5$   & $^3P^o$ 				   &           14.10  &			    17.04\\
	  				 & $^1P^o$ 				   &           17.79  &		       24.83\\
	\hline
	\hline
   
   \end{tabular*}
\end{table}

The harmonic response of Ar$^+$ to intense laser light is also of relevance to experiment. The highest 
harmonics (up to 250 eV) generated by irradiation of Ar by intense laser light have been assigned
to the response of Ar$^+$ \cite{argon+_gibson,argon+_zepf}. A full picture of
HG at high intensities should therefore include ionized species. Recent
photoionization experiments and calculations also provide detailed information on resonances that
may affect HG in Ar$^+$ \cite{photoionization_ar+}.


\section{Theory} 

The TDRM approach is a non-perturbative, {\it ab initio} approach to describe
multielectron atoms in short, intense light pulses. Full details of the method
can be found in \cite{tdrm}.  By propagating the solution of the time-dependent
Schr\"odinger equation on a discrete time mesh of step size $\Delta t$, we can
express its solution, $\Psi_{t_{q+1}}$, at time $t_{q+1}$ in terms of the
solution at the previous time step $t_q$:
\begin{equation} \label{eq:TDSE-sol} (H_m-E)\Psi_{t_{q+1}}=-(H_m+E)\Psi_{t_{q}}.
\end{equation}
Here, $H_m$ is the Hamiltonian at the midpoint of the time interval, and it
contains both the non-relativistic field-free atomic Hamiltonian and the laser
interaction term.  The laser field is assumed to be linearly polarized and
spatially homogeneous and is described by the dipole approximation in the length
form \cite{tdrm_dipole_gauge}. The imaginary energy, $E$, is defined as
$2i/\Delta t$. 

TDRM theory makes use of the $R$-matrix partition of configuration space.
Within an inner region close to the nucleus, full account is
taken of all electron-electron interactions. Outside of this region,
exchange interactions between an ejected electron and those remaining close to
the atomic core can be neglected, and the ejected electron moves only in the
long-range multipole potential of the residual ion and the laser field. 

We evaluate Eq. (\ref{eq:TDSE-sol}) at the boundary of this inner region,
$a_{\mathrm{in}}$, as a matrix equation \cite{tdrm}:
\begin{equation}
  \label{eq:Rmatrix}
  \mathbf{F}(a_{\mathrm{in}})=
  \mathbf{R}(a_{\mathrm{in}})\bar{\mathbf{F}}(a_{\mathrm{in}}) +
  \mathbf{T}(a_{\mathrm{in}}),
\end{equation}
in which the wavefunction, {\bf F}, at the boundary is expressed in terms of
its derivative, $\bar\mathbf{F}$, and an inhomogeneous vector, {\bf T}, which
arises from the right hand side of Eq. (\ref{eq:TDSE-sol}). The matrix {\bf R}
connects the inner and outer region wavefunction at the boundary,
$a_{\mathrm{in}}$.

Given an inner region wavefunction, {\bf R} and {\bf T} are evaluated at the
boundary, and propagated outwards in space to an outer region limit 
at which it can be assumed that the wavefunction has
vanished. There, the wavefunction, $F$, can be set to zero and propagated inwards
to $a_{\mathrm{in}}$. Once $F$ is determined at every boundary point, the full
wavefunction can be extracted from the $R$-matrix equations. We can then iterate
the procedure using Eq. (\ref{eq:TDSE-sol}).

The light radiated by an oscillating dipole is proportional to its acceleration
\cite{acceleration_hhg_sundaram,acceleration_hhg_burnett}.  It is also possible
to express the harmonic spectrum in terms of the dipole velocity or dipole
operators
\cite{length_hhg_eberly,length_hhg_bandarage,dipole_gauge_madsen,velocity_hhg}.
For He, the different expressions have been shown to be both self-consistent
within TDRM, and consistent with the spectrum obtained by the HELIUM approach
\cite{brown_helium, helium_smyth}.  We show only dipole length spectra here. We
have verified that they are consistent with those based on the dipole velocity
form.

The harmonic response of a single atom can be evaluated via the
expectation value of the dipole operator:
\begin{equation}
  \label{eq:dipole_length}
  \mathbf{d}(t)\propto \left< \Psi (t) | \mathbf{z}  | \Psi (t) \right>,
\end{equation}
where $\mathbf{z}$ is the total position operator along the laser polarization axis. The
harmonic spectrum is then proportional to the square of the modulus of the Fourier
transform of $\mathbf{d}(t)$, $|\mathbf{d}(\omega)|^2$. We note that this is a
non-relativistic approximation to the HG process. Inclusion of relativistic
effects, such as spin-orbit coupling, would lead to the population of $M=1$
levels, and this is the subject of ongoing investigation.


\section{Computational setup}
In $R$-matrix theory, Ar$^+$ is described as a state of Ar$^{2+}$ plus an additional
electron. To describe Ar$^{2+}$, we use a set of Hartree-Fock
\{$1s,2s,2p,3s$,$3p$\} orbitals, obtained for the
Ar$^{2+}$ ground state \cite{clementi_atomic_data}.
We obtain the $3s^23p^4$ and $3s3p^5$ eigenstates of Ar$^{2+}$
from configuration-interaction calculations comprising $3s^23p^4$, $3s3p^5$ and $3p^6$.
No pseudo-orbitals are included, since such orbitals may lead to
spectra influenced by unphysical resonances.
A consequence of this basis-set restriction is that the energies of the Ar$^{2+}$ thresholds differ
by 0.2 and 0.4 eV from experiment for the $3s^23p^4$ states and
3 and 7 eV for the $3s3p^5$ states (Tab. \ref{tab:nist-data}).
To assess interference effects, we employ several Ar$^+$ models.
The full Ar$^+$ model contains all five Ar$^{2+}$ thresholds. In the three-state model,
only the $3s^23p^4$ states of Ar$^{2+}$ are retained. We also use Ar$^+$ models
in which only a single state or a pair of states from the $3s^23p^4$ configuration is retained. In these latter models, both
$3s3p^5$ states are always included. 

The inner region has a radius of 15 a.u. which suffices to contain the residual
Ar$^{2+}$  ion. The outer region radius is 600 a.u. The set of continuum
orbitals is built from 60 $B$-splines for each angular momentum of the continuum
electron. The Ar$^+$ basis contains all allowed combinations of Ar$^{2+}$ states
and the set of continuum orbitals up to a total angular momentum $L_{max}=19$.
Convergence testing was carried out up to $L_{max}=23$. The outer region is
divided into sectors of 2 a.u. each containing 35 $B$-splines of order 9 per
channel. The time step in the wavefunction propagation is 0.1 a.u. We use 390-nm
laser pulses, consisting of a 3 cycle $\sin^2$ ramp-on followed by 2 cycles at
peak intensity, 4 $\times$ 10$^{14}$ W/cm$^2$, and a 3 cycle $\sin^2$ ramp-off.
The initial state is the Ar$^+$ ground state with total magnetic quantum number
$M=0$, which would be the dominant non-relativistic Ar$^+$ ground state level
following strong-field ionization of Ar at 390-nm.

\section{Results} 


\begin{figure}[t]
	\centering
	\includegraphics[width=8.7cm]{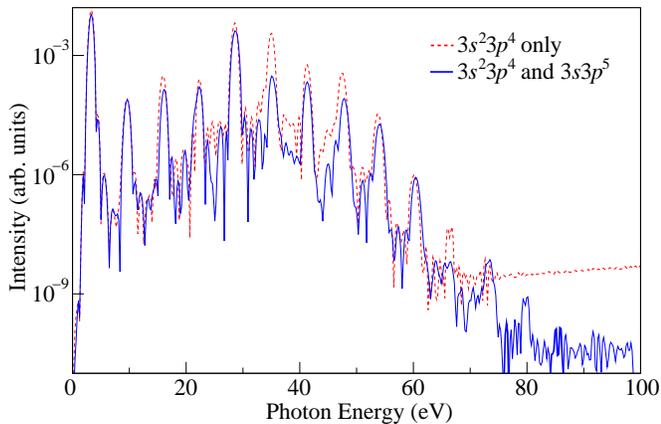}
	\caption{(Color online) Harmonic spectrum of Ar$^+$ produced by an 8-cycle
	  $4\times10^{14}$  Wcm$^{-2}$ laser pulse at 390-nm, as obtained by
            the three-state model (dotted, red line) and by the five-state model
				(solid, blue line).
 \label{fig:3vs5targ}}
\end{figure}

Figure \ref{fig:3vs5targ} shows the harmonic spectra produced from the full and
the three-state model of Ar$^+$. The three-step model \cite{corkum1993} suggests a
harmonic cut-off of 46 eV. Even though the three-step model may not apply at 390-nm,
the harmonic spectra are not inconsistent with this value.

As shown in Fig. \ref{fig:3vs5targ}, inclusion of the $3s3p^5$ thresholds has a
noticeable effect on the yields for the 11th to the 15th harmonic, which are
reduced by up to an order of magnitude. These harmonics coincide with the range
of energies associated with the Rydberg series converging onto the $3s3p^5$
thresholds. We can thus ascribe the differences between the two spectra as
arising from the reaction of multiple electrons to the laser field.  Rydberg
series converging onto $3s3p^5$ effectively describe the excitation of a $3s$
electron, while the main contribution to the full harmonic spectrum involves the
emission of a $3p$ electron. This competition is similar to the one observed for
HG in Ar \cite{brown_prl}.  However, the present study employs a longer
wavelength and involves a five-photon gap between the $3s^23p^4$ and $3s3p^5$
thresholds.

The main focus of the present study is interference between channels associated with the
three low-lying $3s^23p^4$ thresholds. We have therefore calculated harmonic spectra
using Ar$^+$ models in which only selected $3s^23p^4$ thresholds are included.

\begin{figure}[t]
  \centering 
  \includegraphics[width=8.7cm]{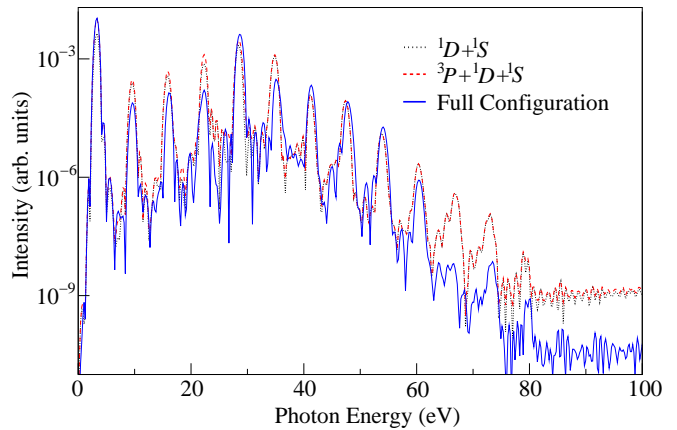}
  \caption{(Color online) Ar$^+$ harmonic spectrum produced by an 8-cycle,
	 $4\times10^{14}$ Wcm$^{-2}$ laser pulse at 390-nm as 
	 calculated by summing contributions of the individual $3s^23p^4$ $^1D$ and $^1S$
	 models (dotted, black line), the individual $3s^22p^4$ $^3P$, $^1D$ and
	 $^1S$ models (dashed red
	 line) and from the full model (solid, blue line).
	 All calculations include both $3s3p^5$ thresholds.
	 \label{fig:sing-full}} 
\end{figure}

Figure \ref{fig:sing-full} compares the harmonic spectra obtained by summing
contributions of the individual $3s^23p^4$ $^1D$ and $^1S$ models, the
individual $3s^22p^4$ $^3P$, $^1D$ and $^1S$ models and the spectrum from the
full model. The individual $3s^23p^4$ models retain both $3s3p^5$ thresholds.
The spectrum is largely dominated by the $^1D$ model, while the $^3P$
model does not contribute significantly, especially in the cutoff region, as
demonstrated by the small difference between the two summed spectra. The largest
difference is seen for the 7th harmonic for which the $^3P$ model
contributes 41\%.  In terms of the three-step mechanism, this suggests that
tunnel ionization leaving Ar$^{2+}$ in the ground state does not significantly
contribute to HG.  In the three-step model for Ar$^+$, the first step should
consider tunnel ionization leaving Ar$^{2+}$ in an excited state.

Figure \ref{fig:sing-full} shows that interference between channels associated
with different $3s^23p^4$ thresholds must be accounted for. The comparison
between the summed spectra and the full model shows differences for the 3rd -7th
harmonic and the 11th harmonic by as much as an order of magnitude, noticeable
shifts in the energy of the 13th and 15th harmonics and reductions by two orders
of magnitude for the 21st and 23rd harmonics.  The accurate determination of
Ar$^+$ harmonic yields thus requires calculations including all $3s^23p^4$
channels simultaneously.

\begin{figure}[t]
	\centering
	\includegraphics[width=8.7cm]{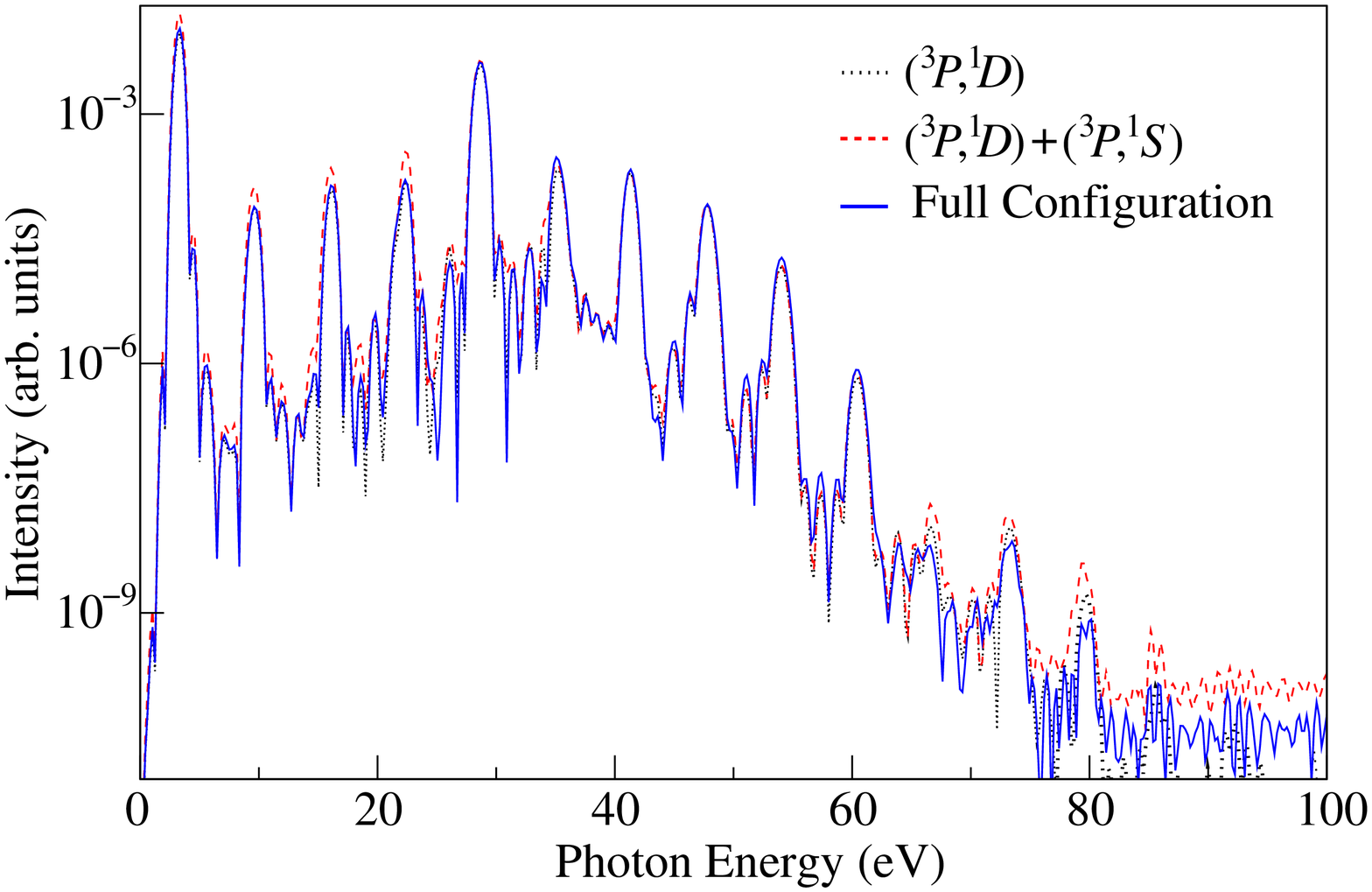}
	\caption{(Color online) Ar$^+$ harmonic spectrum produced by an 8-cycle
	  $4\times10^{14}$ Wcm$^{-2}$ laser pulse  at 390-nm as
	  obtained by the ($^3P$, $^1D$) model (dotted, black line), by
	  summing the spectra obtained by the ($^3P$, $^1D$) model and the
	  ($^3P$, $^1S$) model (dashed, red line) and by
	  the full model (solid, blue line). 
	  \label{fig:doub-full}}
\end{figure}

To assess the interferences, we have performed calculations in which pairs of
$3s^23p^4$ thresholds of Ar$^{2+}$ are retained. Of particular interest are the
spectra in which we retain (a) $3s^23p^4$ $^3P$ and $^1D$  and (b) $3s^23p^4$
$^3P$ and $^1S$. As shown in Fig. \ref{fig:doub-full}, the ($^3P$,$^1D$) model
provides a spectrum with harmonic peaks within 15\% of the full model up to the
15th harmonic, apart from the 11th with a difference of 30\%; the 17th and 19th
peaks differ by 25\%.  Channels associated with the $^3P$ and the $^1D$
thresholds are hence the most important channels for HG. The improvement over
the individual-state models (Fig.  \ref{fig:sing-full}) shows that channels
associated with the $^3P$ threshold are important to HG beyond the $^3P$
threshold, even though their direct contribution to these harmonics is
relatively minor.

Addition of the harmonic spectra obtained from the ($^3P$,$^1D$) and the
($^3P$,$^1S$) model shows poor agreement for harmonics below the Ar$^+$
ionization threshold as the Rydberg series leading up to the $^3P$ threshold
contributes twice. However, beyond this threshold, the agreement with the full
model is very good, with harmonic peaks differing by up to only 6\% for the 9th
to the 19th, apart from differences of 15\% and 19\% for the 11th and 17th
harmonics.  This comparison again indicates that HG due to channels associated with
the $^3P$ threshold primarily occurs below the ionization threshold. Above this
threshold, HG from these channels is negligible.

The importance of the $^3P$ threshold for an accurate harmonic spectrum can be
understood by considering the atomic structure. The first step within the
three-step model is emission of a single electron.  For an accurate emission
rate, the atomic structure up to the threshold must be described well.
Inclusion of the $^3P$ threshold is necessary to account for the Rydberg series
leading up to the first ionization threshold. This series also affects the
position of low-lying states such as $3s3p^6$ and of low-lying members of
Rydberg series leading up to other thresholds. Therefore, when HG originates
from an excited threshold, atomic structure associated with the lower thresholds
must still be accounted for.

Finally, we compare the efficiency of HG in Ar$^+$ with that of other atoms.
Ar$^+$ has a slightly higher ionization potential, 27.6 eV, than He, 24.6 eV. On
the other hand, Ar$^+$ is a larger ion, and may therefore provide a greater
harmonic response \cite{argon+_gibson}.  We also consider the Ne$^+$ ion as it
also is an ionized noble-gas atom, but with a substantially higher ionization
potential, 40.96 eV.

\begin{figure}[b]
	\centering
	\includegraphics[width=8.7cm]{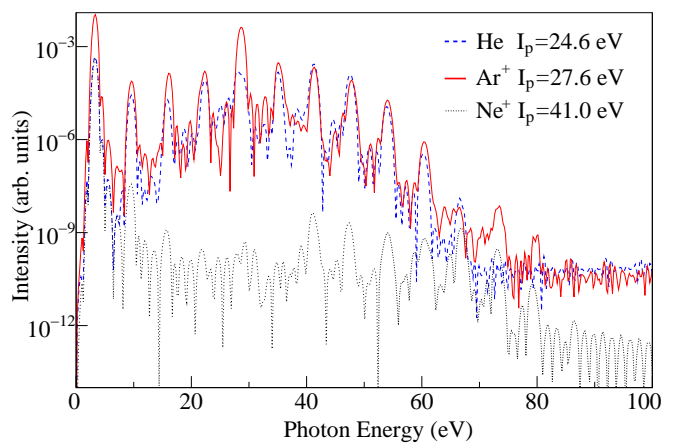}
	\caption{(Color online) The harmonic spectrum produced by a
	  $4\times10^{14}$ Wcm$^{-2}$ 390nm laser pulse interacting with Ar$^+$ (solid, red
	  line), Ne$^+$ (dotted, black line) and He (dashed, blue line). The Ne$^+$
	  spectrum is orders of magnitude lower than the Ar$^+$ and He spectra, but the Ar$^+$
	  yield is slightly larger than the He, despite its higher ionization
	  potential.\label{fig:hel-neo-arg}}
\end{figure}

Figure \ref{fig:hel-neo-arg} shows the harmonic spectrum of Ar$^+$, He and
Ne$^+$ for the laser pulse used throughout. For He, we use the
6P-model described in our earlier work \cite{brown_helium}, whereas for Ne$^+$ we include the
three $2s^22p^4$ and the two $2s2p^5$ thresholds with the Ne$^{2+}$
states generated following a similar procedure as outlined above for Ar$^{2+}$.
Figure \ref{fig:hel-neo-arg} shows expected behaviour for Ne$^+$ and its higher
ionization potential, with a harmonic yield several orders of magnitude smaller
than that of He and Ar$^+$. However, the figure also shows that, apart from the 13th and
15th harmonics, the harmonic yield from Ar$^+$ is consistently higher than
that from He, despite its larger ionization potential. Thus Ar$^+$ is indeed an
efficient ion for HG.


\section{Conclusion} 

We have applied TDRM theory to determine harmonic yields for ground-state Ar$^+$
with $M=0$ and assess in detail the role of the various closely spaced
ionization thresholds.  The dominant contribution to the harmonic yield is
associated with the first excited threshold instead of the lowest ionization
threshold.  This lowest threshold must still be accounted for, as it affects the
atomic structure leading up to excited thresholds.  The $3s3p^5$ thresholds affect the harmonic
yield significantly less, but need to be taken into account for harmonics with
energies close to these thresholds.  Overall, the harmonic yield for Ar$^+$ is
generally larger than the yield for He, even though Ar$^+$ has a larger
ionization potential.

In our discussion of the HG process we have frequently made reference to the
three-step model, even though it may not be applicable 
this wavelength/intensity regime. However, the present calculation demonstrates
that interaction between different channels is important for an accurate
description of HG. Since the energy separation between the $^3P$ and $^1D$
thresholds is comparable to 800-nm photon energies we would expect these
interactions to be important at these wavelengths as well. In these $M=0$
calculations, the highest harmonics ($E > I_p$) are strongly associated with
the Ar$^{2+}$ $^1D$ channels. These channels would not be strongly {\it
disfavored } at 800-nm because of the small energy gap between the $^3P$ and
$^1D$ thresholds. We would therefore expect the $^1D$ channels to remain
important for HG in Ar$^+$ with $M=0$ at 800-nm. Hence, application of the
three-step model for Ar$^+$ with $M=0$ should account for multichannel
interactions.

The harmonic spectra obtained in the present study demonstrate a significant
influence from interference associated with different ionization thresholds.
The use of multielectron codes, such as TDRM, is therefore essential for
systems with excited thresholds just above the lowest ionization threshold.
Atomic systems, such as Ar$^+$, may be suitable for developing understanding of
how multichannel interactions affect HG. Further experimental studies of HG in
Ar$^+$ would thus be very interesting. We have only addressed the
non-relativistic case of $M=0$ here, but aim to extend the work to include the
effect of $M=1$ in our calculations. We also aim to extend these studies to
longer wavelengths, which will enable more detailed comparisons with experiment.
The recently developed $R$-matrix with time-dependence approach \cite{RMT,RMT2}
will be of major benefit for this extension.

ACB acknowledges support from DEL (NI). HWH is supported by EPSRC under
grant number G/055416/1.

\bibliography{mybib}
\end{document}